\documentclass[twocolumn]{aastex63}
\usepackage{amsmath}
\usepackage{multirow}
\usepackage{booktabs}
\usepackage{xcolor}

\shorttitle{Galactic Latitude Dependence of FRB Sky Distribution}
\shortauthors{Josephy et al.}
\graphicspath{{./}{figures/}}
\date{\today}

\begin{document}

\title{No Evidence for Galactic Latitude Dependence of the Fast Radio Burst Sky Distribution}
\correspondingauthor{A. Josephy}
\email{alexander.josephy@mail.mcgill.ca}

\author[0000-0003-3059-6223]{A.~Josephy}
  \affiliation{Department of Physics, McGill University, 3600 rue University, Montr\'eal, QC H3A 2T8, Canada}
  \affiliation{McGill Space Institute, McGill University, 3550 rue University, Montr\'eal, QC H3A 2A7, Canada}
\author[0000-0002-3426-7606]{P.~Chawla}
  \affiliation{Department of Physics, McGill University, 3600 rue University, Montr\'eal, QC H3A 2T8, Canada}
  \affiliation{McGill Space Institute, McGill University, 3550 rue University, Montr\'eal, QC H3A 2A7, Canada}
\author[0000-0002-8376-1563]{A.~P.~Curtin}
  \affiliation{Department of Physics, McGill University, 3600 rue University, Montr\'eal, QC H3A 2T8, Canada}
  \affiliation{McGill Space Institute, McGill University, 3550 rue University, Montr\'eal, QC H3A 2A7, Canada}  
\author[0000-0001-9345-0307]{V.~M.~Kaspi}
  \affiliation{Department of Physics, McGill University, 3600 rue University, Montr\'eal, QC H3A 2T8, Canada}
  \affiliation{McGill Space Institute, McGill University, 3550 rue University, Montr\'eal, QC H3A 2A7, Canada}
\author[0000-0002-3615-3514]{M.~Bhardwaj}
  \affiliation{Department of Physics, McGill University, 3600 rue University, Montr\'eal, QC H3A 2T8, Canada}
  \affiliation{McGill Space Institute, McGill University, 3550 rue University, Montr\'eal, QC H3A 2A7, Canada}
\author[0000-0001-8537-9299]{P.~J.~Boyle}
  \affiliation{Department of Physics, McGill University, 3600 rue University, Montr\'eal, QC H3A 2T8, Canada}
  \affiliation{McGill Space Institute, McGill University, 3550 rue University, Montr\'eal, QC H3A 2A7, Canada}
\author[0000-0002-1800-8233]{C.~Brar}
  \affiliation{Department of Physics, McGill University, 3600 rue University, Montr\'eal, QC H3A 2T8, Canada}
  \affiliation{McGill Space Institute, McGill University, 3550 rue University, Montr\'eal, QC H3A 2A7, Canada}
\author[0000-0003-2047-5276]{T.~Cassanelli}
  \affiliation{Dunlap Institute for Astronomy \& Astrophysics, University of Toronto, 50 St.~George Street, Toronto, ON M5S 3H4, Canada}
  \affiliation{David A.~Dunlap Department of Astronomy \& Astrophysics, University of Toronto, 50 St.~George Street, Toronto, ON M5S 3H4, Canada}
\author[0000-0001-8384-5049]{E.~Fonseca}
  \affiliation{Department of Physics, McGill University, 3600 rue University, Montr\'eal, QC H3A 2T8, Canada}
  \affiliation{McGill Space Institute, McGill University, 3550 rue University, Montr\'eal, QC H3A 2A7, Canada}
\author[0000-0002-3382-9558]{B.~M.~Gaensler}
  \affiliation{Dunlap Institute for Astronomy \& Astrophysics, University of Toronto, 50 St.~George Street, Toronto, ON M5S 3H4, Canada}
  \affiliation{David A.~Dunlap Department of Astronomy \& Astrophysics, University of Toronto, 50 St.~George Street, Toronto, ON M5S 3H4, Canada}
\author[0000-0002-4209-7408]{C.~Leung}
  \affiliation{MIT Kavli Institute for Astrophysics and Space Research, Massachusetts Institute of Technology, 77 Massachusetts Ave, Cambridge, MA 02139, USA}
  \affiliation{Department of Physics, Massachusetts Institute of Technology, 77 Massachusetts Ave, Cambridge, MA 02139, USA}
\author[0000-0001-7453-4273]{H.-H.~Lin}
  \affiliation{Canadian Institute for Theoretical Astrophysics, 60 St.~George Street, Toronto, ON M5S 3H8, Canada}
\author[0000-0002-4279-6946]{K.~W.~Masui}
  \affiliation{MIT Kavli Institute for Astrophysics and Space Research, Massachusetts Institute of Technology, 77 Massachusetts Ave, Cambridge, MA 02139, USA}
  \affiliation{Department of Physics, Massachusetts Institute of Technology, 77 Massachusetts Ave, Cambridge, MA 02139, USA}
\author[0000-0001-7348-6900]{R.~Mckinven}
  \affiliation{Dunlap Institute for Astronomy \& Astrophysics, University of Toronto, 50 St.~George Street, Toronto, ON M5S 3H4, Canada}
  \affiliation{David A.~Dunlap Department of Astronomy \& Astrophysics, University of Toronto, 50 St.~George Street, Toronto, ON M5S 3H4, Canada}
\author[0000-0002-0772-9326]{J.~Mena-Parra}
  \affiliation{MIT Kavli Institute for Astrophysics and Space Research, Massachusetts Institute of Technology, 77 Massachusetts Ave, Cambridge, MA 02139, USA}
\author[0000-0002-2551-7554]{D.~Michilli}
  \affiliation{Department of Physics, McGill University, 3600 rue University, Montr\'eal, QC H3A 2T8, Canada}
  \affiliation{McGill Space Institute, McGill University, 3550 rue University, Montr\'eal, QC H3A 2A7, Canada}
\author[0000-0002-3616-5160]{C.~Ng}
  \affiliation{Dunlap Institute for Astronomy \& Astrophysics, University of Toronto, 50 St.~George Street, Toronto, ON M5S 3H4, Canada}
\author[0000-0002-4795-697X]{Z.~Pleunis}
  \affiliation{Department of Physics, McGill University, 3600 rue University, Montr\'eal, QC H3A 2T8, Canada}
  \affiliation{McGill Space Institute, McGill University, 3550 rue University, Montr\'eal, QC H3A 2A7, Canada}
\author[0000-0001-7694-6650]{M.~Rafiei-Ravandi}
  \affiliation{Perimeter Institute for Theoretical Physics, 31 Caroline Street N, Waterloo, ON N25 2YL, Canada}
  \affiliation{Department of Physics and Astronomy, University of Waterloo, Waterloo, ON N2L 3G1, Canada}
\author[0000-0003-1842-6096]{M.~Rahman}
  \affiliation{Dunlap Institute for Astronomy \& Astrophysics, University of Toronto, 50 St.~George Street, Toronto, ON M5S 3H4, Canada}
  \affiliation{Sidrat Research, PO Box 73527 RPO Wychwood, Toronto, ON M6C 4A7, Canada}
\author[0000-0001-5504-229X]{P.~Sanghavi}
  \affiliation{Lane Department of Computer Science and Electrical Engineering, 1220 Evansdale Drive, PO Box 6109 Morgantown, WV 26506, USA}
  \affiliation{Center for Gravitational Waves and Cosmology, West Virginia University, Chestnut Ridge Research Building, Morgantown, WV 26505, USA}
\author[0000-0002-7374-7119]{P.~Scholz}
  \affiliation{Dunlap Institute for Astronomy \& Astrophysics, University of Toronto, 50 St.~George Street, Toronto, ON M5S 3H4, Canada}
\author[0000-0002-6823-2073]{K.~Shin}
  \affiliation{MIT Kavli Institute for Astrophysics and Space Research, Massachusetts Institute of Technology, 77 Massachusetts Ave, Cambridge, MA 02139, USA}
  \affiliation{Department of Physics, Massachusetts Institute of Technology, 77 Massachusetts Ave, Cambridge, MA 02139, USA}
\author[0000-0002-2088-3125]{K.~M.~Smith}
  \affiliation{Perimeter Institute for Theoretical Physics, 31 Caroline Street N, Waterloo, ON N25 2YL, Canada}
\author[0000-0001-9784-8670]{I.~H.~Stairs}
  \affiliation{Department of Physics and Astronomy, University of British Columbia, 6224 Agricultural Road, Vancouver, BC V6T 1Z1 Canada}
\author[0000-0003-2548-2926]{S.~P.~Tendulkar}
  \affiliation{Department of Astronomy and Astrophysics, Tata Institute of Fundamental Research, Mumbai, 400005, India}
  \affiliation{National Centre for Radio Astrophysics, Post Bag 3, Ganeshkhind, Pune, 411007, India}
\author[0000-0001-8278-1936]{A.~V.~Zwaniga}
  \affiliation{Department of Physics, McGill University, 3600 rue University, Montr\'eal, QC H3A 2T8, Canada}

\begin{abstract}
We investigate whether the sky rate of Fast Radio Bursts depends on Galactic latitude using the first catalog of Fast Radio Bursts (FRBs) detected by the Canadian Hydrogen Intensity Mapping Experiment Fast Radio Burst (CHIME/FRB) Project.  We first select CHIME/FRB events above a specified sensitivity threshold in consideration of the radiometer equation, and then compare these detections with the expected cumulative time-weighted exposure using Anderson-Darling and Kolmogrov-Smirnov tests.  These tests are consistent with the null hypothesis that FRBs are distributed without Galactic latitude dependence ($p$-values distributed from 0.05 to 0.99, depending on completeness threshold). Additionally, we compare rates in intermediate latitudes ($|b| < 15^\circ$) with high latitudes using a Bayesian framework, treating the question as a biased coin-flipping experiment-- again for a range of completeness thresholds. In these tests the isotropic model is significantly favored (Bayes factors ranging from 3.3 to 14.2). Our results are consistent with FRBs originating from  an isotropic population of extragalactic sources.
\end{abstract}
\keywords{Radio transient sources (2008); High energy astrophysics (739)}
\section{Introduction}\label{sec:introduction}

Fast Radio Bursts (FRBs) are  $\mu$s$-$ms duration radio bursts with dispersion measures (DMs) consistent with origins far outside our Milky Way galaxy.  Thirteen FRB sky localizations \citep{tbc+17,clw+17,bdp+19,rcd+19,mpm+20,mnh+20,hps+20} have identified host galaxies, most at cosmological distances, but which are of diverse types that have not yet clarified the nature of these transient sources.  Some FRBs repeat \citep{ssh+16a,abb+19b,abb+19c,fab+20,kso+19}, and repeating sources exhibit bursts having properties that differ temporally and spectrally from those of apparent non-repeaters \citep{ssh+16a,abb+19c,hss+19,fab+20,p++21}. Hence, there may be more than one class of bursting radio sources comprising FRBs.

One puzzling observational feature of FRBs early on in their study was evidence for a Galactic latitude dependence of their sky distribution.  \citet{psj+14} reported that FRBs avoid the Galactic plane at a 99\% confidence-level based on four events detected at high latitudes and a lack of detections in an intermediate latitude FRB survey ($-15^{\circ}<b<15^{\circ})$, with both surveys done using the 64-m Murriyang telescope\footnote{Formerly known as the Parkes Radio Telescope.} at the Parkes Observatory. Further observations reporting an additional two bursts  at high latitude supported this claim, arguing for a $\sim3\sigma$ dearth at low latitudes \citep{bb14}. \citet{cpk+16} found, following the detection of an additional five events, that the intermediate and high latitude rates are inconsistent with 97.5\% confidence.  However, \citet{cpo16} used the same data set but different statistical techniques to argue against any latitude dependence.  \citet{rlb+16} also argued against a dependence, attributing the perceived deficit to initially overestimated FRB rates.  Most recently, based on a sample of fifteen FRBs, \citet{bkb+18} concluded that any Galactic latitude effect has low ($<2\sigma$) significance, and that past claims could be explained by  either small samples of FRBs, hence poor statistics, or inaccuracies in the modeling of Galactic dispersion and scattering effects.

Meanwhile, \citet{mj15} suggested that diffractive scintillation could effectively boost the apparent number of high-latitude sources even for an underlying isotropic distribution.  However this requires a very steep FRB flux density distribution.  The scarcity of Arecibo FRB discoveries provided evidence against this distribution, given the Arecibo telescope's high sensitivity \citep{ssh+16b}.  While unlikely given the high DMs of the Parkes detections, another  generic reason for a low-level Galactic latitude dependence could be the presence of a relatively small fraction of Galactic sources within the FRB population at high latitudes, on top of a largely isotropic distribution, with the former misidentified as extragalactic due to inaccuracies in Galactic DM-distribution models.

The Canadian Hydrogen Intensity Mapping Experiment (CHIME\footnote{\url{https://chime-experiment.ca}}) Fast Radio Burst project \citep[hereafter CHIME/FRB;][]{abb+18} is well suited to searching for Galactic latitude dependence of the FRB sky distribution because of the system's high event rate and uniform, repeated coverage of the sky northward of declination $-11^{\circ}$.  Operating since mid-2018 in the 400--800 MHz band, the first CHIME/FRB catalog \citep{chimefrbcatalog1}, hereafter Catalog 1, reports on 535 events detected with a single search pipeline, hence uniform selection biases.  Here we report on our search for any Galactic latitude dependence of the FRB sky distribution using the Catalog 1 sample.

 The paper's outline is as follows. In \S\ref{sec:observations} the observations are discussed, namely the exposure maps and the sample of CHIME/FRB events included in the analysis. Sensitivity correction methods are presented in \S\ref{sec:sensitivity}. Results from two classes of statistical tests are given in  \S\ref{sec:tests} and outcomes are discussed in \S\ref{sec:discussion}. Conclusions are provided in \S\ref{sec:conclusions}.\\

\section{Observations}\label{sec:observations}
\begin{figure}[htb]
    \centering
    \includegraphics[width=0.95\linewidth]{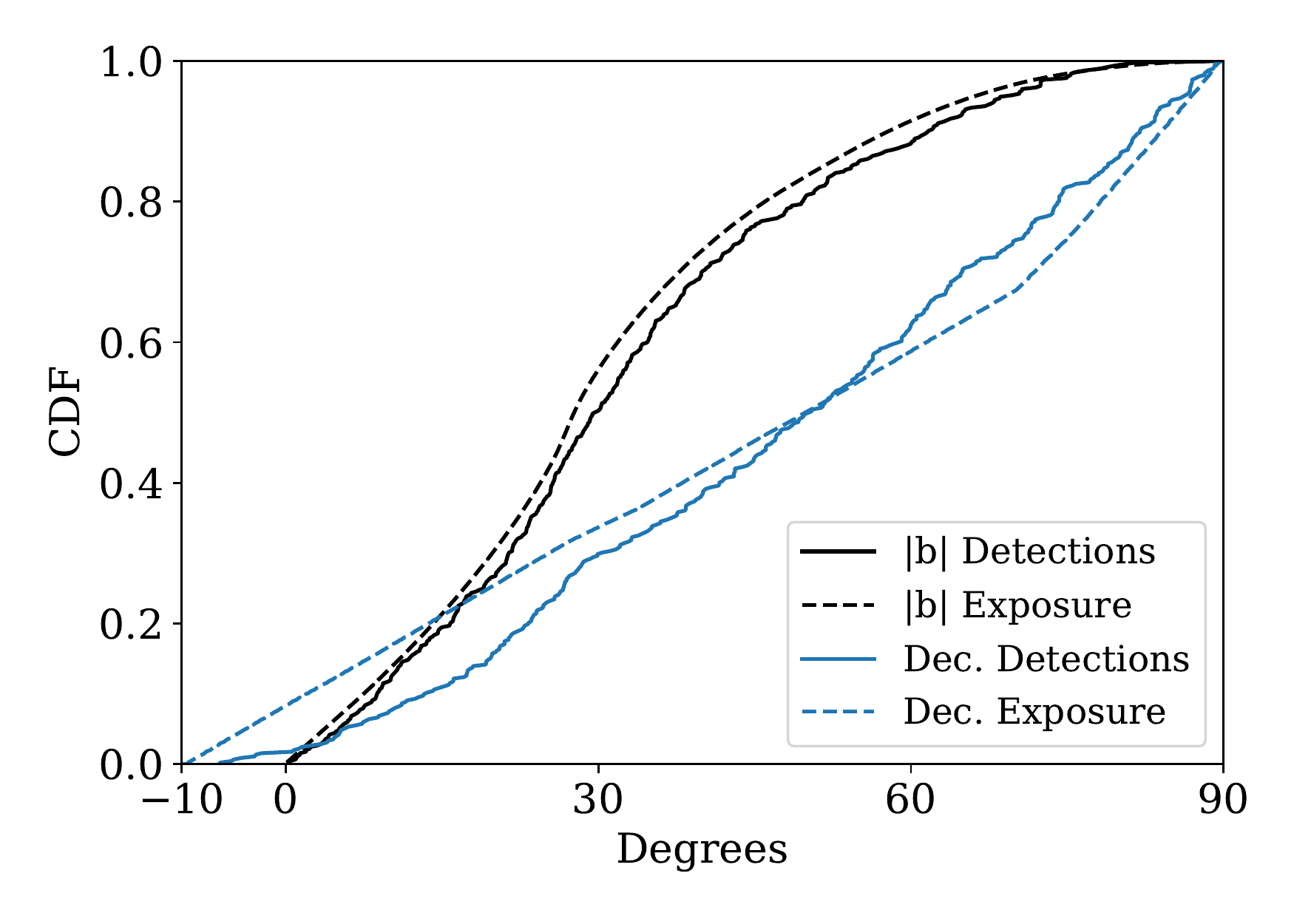}
    \caption{Cumulative exposure (dashed) and cumulative source positions (solid) as a function of absolute Galactic latitude (black) and as a function of declination (blue). Note the slope increase above declination ${\sim}70^\circ$ is due to doubled exposure from the secondary ``lower" transit. Only the cuts described in~\S\ref{sec:observations} have been applied to the original Catalog 1 sample of bursts.}
    \label{fig:exposure}
\end{figure}

 The primary observational inputs taken from Catalog 1 are the exposure map, source positions, and detection signal-to-noise ratios (S/N; used for completeness cuts). Exposure maps are provided in a HEALPix\footnote{\url{https://healpix.sourceforge.io}} format \citep{ghb+05} and  span from 2018 August 28 to 2019 July 1. Notably, the integration range does not cover all bursts in Catalog 1, as the reported exposure is meant to capture relatively uniform observing conditions (e.g. singular beam configuration and stable sensitivity). The standard of uniformity is crucial for this analysis, so we excise events accordingly. The first thirteen bursts \citep{abb+19a} were detected during a precommissioning phase with a considerably different beam configuration. A further 26 events occurred on days with either abnormally low sensitivity, days with disruptive hardware/software upgrades, or days with significant pipeline issues. Three events are removed for being detected in the far side-lobes of the telescope, as obtaining their positions requires methods beyond the scope of Catalog 1. Finally, we include only a single burst from each of the eighteen repeating FRB sources in Catalog 1, taking the event with the highest S/N (this choice is to accommodate subsequent completeness cuts, see \S\ref{sec:cuts}). The final sample used in this paper includes 453 bursts. The exposure products provided by Catalog 1 have been defined at the FWHM of the synthesized beams at 600~MHz, which necessitates a high map resolution to capture  gaps in the exposure between beams. Such a resolution is not required for this work, so we downsample the maps to have pixel area $\simeq47$ square arcminutes and use the same resolution when generating the sensitivity maps used for completeness cuts (see \S\ref{sec:sensitivity}). The beam configuration for this exposure spans 120$^{\circ}$ North/South in elevation angle centered at zenith, so sources with a declination greater than ${\sim}+70^\circ$ are observed twice a day. Because of this, separate exposure maps exist for the upper and lower transits. Cumulative exposure and source positions along Galactic latitude and declination coordinates are shown in Figure~\ref{fig:exposure} (with only the cuts described in this section applied to the data). The sum total of the exposure is ${\simeq}6.6\times10^5$ hours$\,\cdot\,$deg$^2$. Note that the analysis presented in this paper uses the S/N as the relevant quantity rather than fluence when making completeness cuts.  The current localization uncertainties lead to large uncertainties in the fluence measurements, which are treated as lower limits as bursts are assumed to occur in the center of the formed beams for fluence calculations.  As part of Catalog 1, a fluence completeness threshold is estimated for each burst via Monte Carlo simulation. These methods aim to capture a range of sensitivity variation and are described in detail by \citet{jcf+19}. The reported completeness thresholds are 95\% one-sided confidence intervals, and the median value of these thresholds is $\sim$5 Jy$\,\cdot\,$ms.  The triggering threshold of the real time pipeline is a S/N of 9, and we adopt this value in the subsequent analysis for burst detectability. As for the DM triggering criteria, CHIME/FRB will label an event as extragalactic when the DM is larger than either of the two main Galactic free-electron models estimates (NE2001: \citealt{ne2001}; YMW16: \citealt{ymw17}), using the formed beam center as the line-of-sight. Note that this is different from initial plans \citep{abb+18}. The change was made early in the precommissioning phase and has the advantage of being simpler, more generous, and without reference to mixed uncertainty estimations.

\section{Sensitivity Maps}\label{sec:sensitivity}
 From Figure~\ref{fig:exposure}, a na\"ive comparison of the cumulative CHIME/FRB events against exposure does not qualitatively suggest a Galactic latitude dependence. However, there is a  noticeable deviation from isotropy when declination is taken as the coordinate of interest. This deviation is not a surprise given the significant sensitivity fall-off in the primary beam with zenith angle  (see Figure~\ref{fig:beam-profile}). While there is an interesting astrophysical motivation for examining Galactic latitude rate dependence (e.g., evaluating  Galactic free-electron density models), a rate dependence with declination must be instrumental in the case of CHIME/FRB. Conducting the statistical tests against both Galactic latitude and declination coordinates in parallel serves as an important cross-check that instrumental effects are not corrupting the main focus of this study. To properly compare exposure and detections, we aim to filter out events below a completeness threshold.  In the context of this work, we say a burst is ``above completeness" if we expect it to be reliably detected throughout the given field of view. We can make this comparison with a relative sensitivity map, which we model using a reduced version of the single-pulse form of the radiometer equation \citep[e.g.,][]{cm03}:

\begin{equation}\label{eq:radiometer}
    S_{min} \propto \frac{T_{sys}}{G \sqrt{\Delta \nu}} \frac{\sqrt{w_b}}{w_i}.
\end{equation}

Here $S_{min}$ is the minimum detectable flux density, $T_{sys}$ is the system temperature, $G$ is the telescope gain, $\Delta\nu$ is the observing bandwidth, $w_b$ is the broadened pulse width, and $w_i$ is the intrinsic pulse width. Note that factors common to the CHIME/FRB instrument have been dropped here, namely the number of polarizations, S/N threshold, and scalar efficiency (i.e., digitization loss). As the scope of this work is limited to investigating a differential rate in the CHIME/FRB sample, we can afford to operate in relative terms, which simplifies the analysis substantially. The system temperature, forward gain, effective bandwidth, and pulse-broadening terms are treated separately. Each term leads to a relative sensitivity map in HEALPix format and a composite map is formed through simple multiplicative combination. Just as separate exposure maps exist for the upper and lower transit, we produce two sensitivity maps to capture the double-valued gain. The components of the composite map are described below.

\subsection{System Temperature}
Sensitivity is linearly proportional to the system temperature, which we take to be $T_{sys} = T_{sky} + 50$K using the receiver temperature from \cite{abb+18}.  We estimate sky temperature starting with the destriped, but not desourced, Haslam all-sky continuum map at 408~MHz \citep{rdb+15}. The map is dominated by synchrotron radiation, which is well described by a power-law frequency dependence. We use the spectral index map from the Wilkinson Microwave Anisotropy Probe \citep[WMAP;][]{bhh+03} in conjunction with the Haslam map to compute a band-averaged sky temperature. The mean temperatures for intermediate ($|b| < 15^\circ$) and high Galactic latitudes are 21~K and 10~K respectively,  resulting in a ${\sim}18$\% mean difference in sensitivity.  Note that, for the frequencies (1.22$-$1.52~GHz) and field of view relevant to \citet{psj+14}, the differential sensitivity due to sky temperature was much less significant.

\subsection{Gain}
\begin{figure}[htb]
    \centering
    \includegraphics[width=0.95\linewidth]{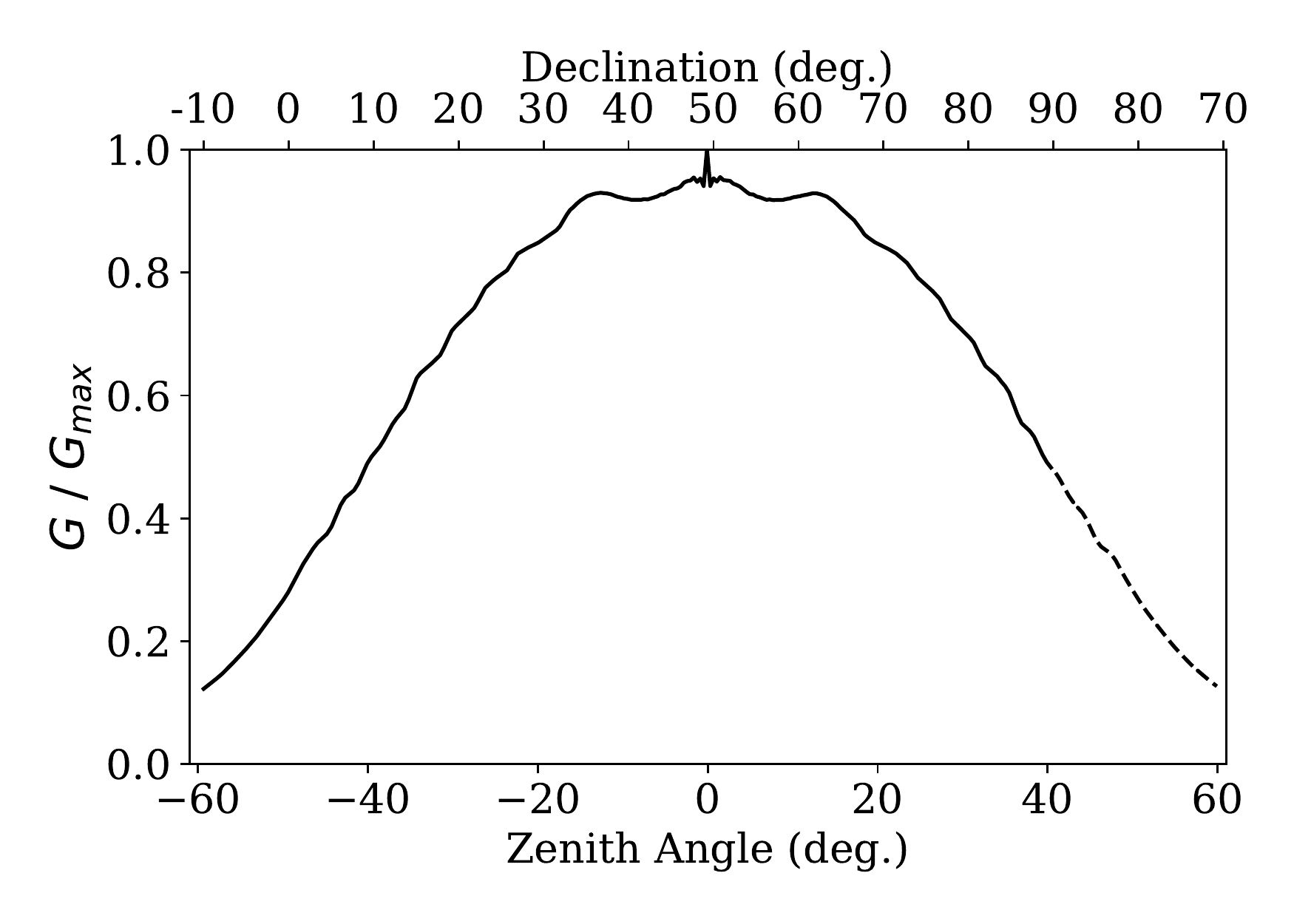}
    \caption{Normalized band-averaged gain (intensity beam response) along meridian. Sensitivity obtained with a composite beam model which includes primary and formed beam effects. Each point represents an average sensitivity within a synthesized beam's 600-MHz FWHM. The dashed portion of the profile shows the lower transit. The sensitivity spike at zenith is a consequence of the beamforming technique \citep{nvp+17}.}
    \label{fig:beam-profile}
\end{figure}

The sensitivity varies dramatically across the field of view. The most apparent effect is a roll-off with increasing zenith angle.  We use a composite beam model that includes a frequency-dependent description of both the primary beam and the synthesized beam pattern. The synthesized component is described by \citet{nvp+17}, while the most recent description of the primary component is given by \citet{chimefrbcatalog1}. The model computes a normalized sensitivity after specifying a frequency and position.  To generate a senstivity map, we compute a beam and band averaged sensitivity for a fine gridding along the meridian (see Figure~\ref{fig:beam-profile}). Averaging for the formed beams is done within the FWHM at 600 MHz.

\subsection{Bandwidth}
\begin{figure}[htb]
    \centering
    \includegraphics[width=0.95\linewidth]{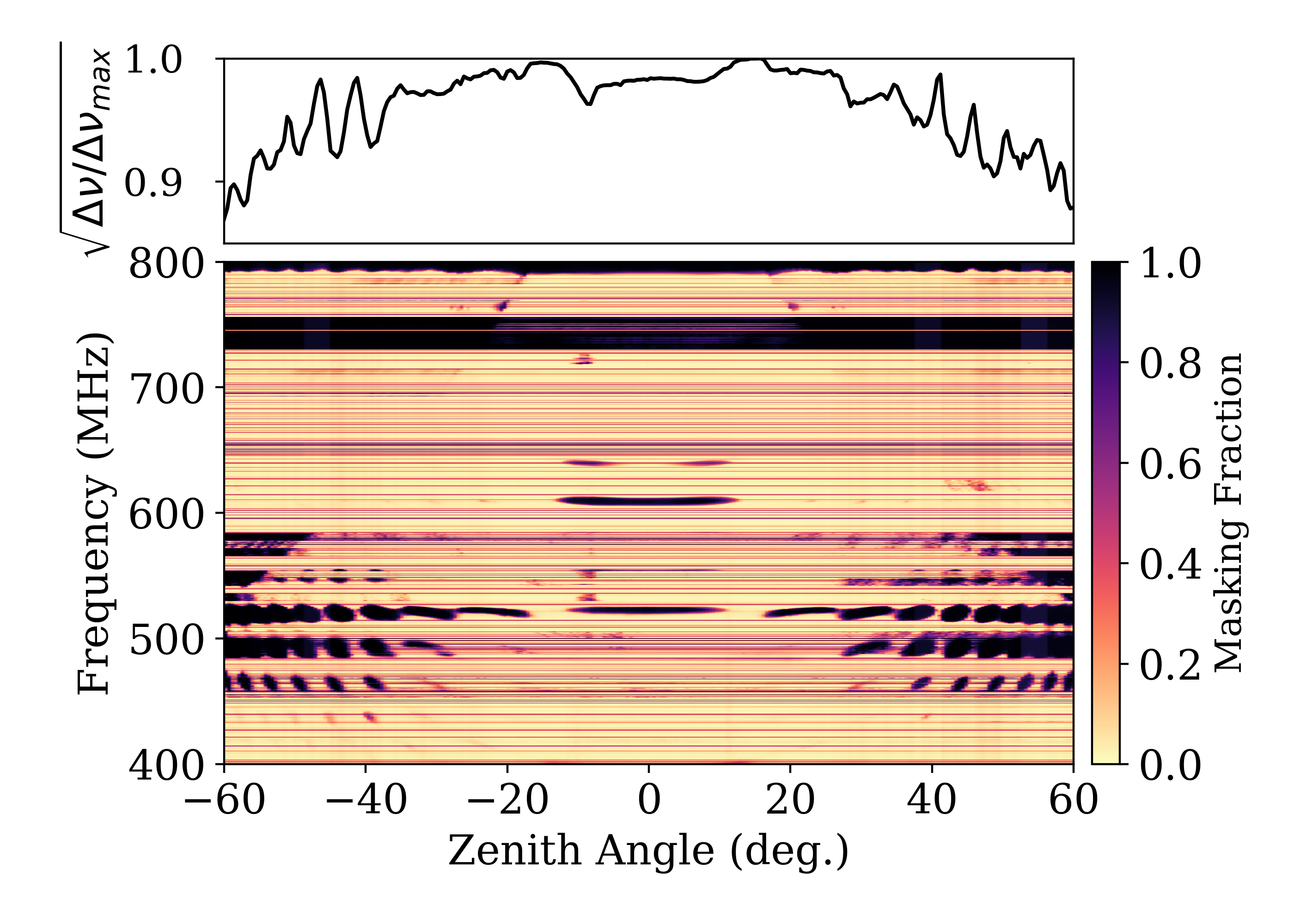}
    \caption{RFI masking fraction as a function of zenith angle and frequency. Masking metrics are averaged over the four beams in the East-West direction. The top panel shows the square-root of the normalized bandwidth, which translates to a relative sensitivity.}
    \label{fig:rfi-mask}
\end{figure}

Minimum detectable fluence is inversely proportional to the square root of the bandwidth. The total available band fluctuates over time due to correlator cluster health \citep{drv+20}, and is further reduced by realtime masking of radio frequency interference (RFI).  To quantify any spatial bandwidth dependence, we recorded metrics of the masking fraction as a function of frequency (downsampled to 1024 channels), on a per-beam basis, downsampled from 1~ms to 60~s. Since we are building towards a singular sensitivity map that reflects bandwidth losses due to persistent RFI, we average across the four East-West beam columns and average in time across several metric collection runs from 2020 March 22 to 2020 July 30 (see Figure~\ref{fig:rfi-mask}). Correlator related fluctuations are included in these metrics, but affect all formed beams equally and are subdominant to RFI masking, so we do not pursue an independent characterization of the correlator health over time. The resulting mask has a complicated spatial structure with an increase in masking towards the horizon. The differential effective bandwidth amounts to a relative sensitivity loss of roughly 10\%. The ability to stream and aggregate the RFI metrics was implemented after the 2019 July 1 cutoff date for the exposure.  To investigate whether or not the RFI environment had changed significantly, we visually compared the time-averaged masks with the per-event masks, which are saved along with intensity data for each event. The enhanced masking from 450~MHz to 600~MHz near the horizons and the midband hot-spot that is seen in Fig.~\ref{fig:rfi-mask} are also apparent in the per-event masks after averaging event masks by beam row, suggesting that direction dependent RFI conditions have not changed considerably.

\subsection{Pulse Broadening}\label{sec:pulse-broadening}
Unlike the preceding sensitivity corrections, effects due to pulse-broadening are difficult to apply uniformly as they vary on a pulse-to-pulse basis. A narrow pulse originating relatively nearby will be scatter-broadened by the Galaxy to a larger fractional degree as compared to a distant and intrinsically broad pulse.  We take the broadened width as the quadrature sum

\begin{equation}\label{eq:pulse-broadening}
    w_b = \sqrt{w_i^2 + t_{samp}^2 + t_{chan}^2 + \tau_{sc}^2},
\end{equation}
where $w_i$ is the intrinsic width, $t_{samp}$ is the sampling time (0.983~ms), $\tau_{sc}$ is the scattering timescale, and $t_{chan}$ is the intra-channel dispersion smearing,

\begin{equation}
    t_{chan} = 8.3 \mu\mathrm{s} \left(\frac{\Delta\nu_{chan}}{\mathrm{MHz}}\right)
    \left(\frac{\nu}{\mathrm{GHz}}\right)^{-3}\left(\frac{\mathrm{DM}}{\mathrm{pc\cdot cm^{-3}}}\right).
\end{equation}
Here $\Delta\nu_{chan}$ is the channel bandwidth for CHIME/FRB (24.4~kHz) and $\nu$ is the channel frequency. For reference, the intra-channel smearing at 600~MHz for DM $= 500\ {\mathrm{pc\cdot cm^{-3}}}$ is ${\sim}$0.5~ms for CHIME/FRB. Galactic scattering timescales are estimated with electron density models (NE2001: \citealt{ne2001}; YMW16: \citealt{ymw17}) and we use a $\tau_{sc} \propto \nu^{-4}$ relationship to appropriately scale from 1~GHz \citep{bcc+04}. We simplify Galactic scattering as originating from a screen located 25 kpc away and include the geometrical lever-arm effect \citep[][]{lkmj13}

\begin{equation}
    \tau_{sc} = \tau_{max}\cdot 4f(1-f);\quad f = \frac{25\mathrm{kpc}}{D_{L}},
\end{equation}
where $\tau_{max}$ is the maximum scattering timescale achieved while integrating the models out to 25~kpc and $D_L$ is the luminosity distance to the FRB source.  Altogether, a relative sensitivity map due to pulse-broadening depends on $w_i$, $DM$, $D_L$, and $\nu$.  Since such a map is burst dependent, we construct an averaged map using Monte Carlo methods, taking a log-normal fit of the intrinsic widths from Catalog 1 to specify a width distribution. For a distance we take 100~Mpc for all bursts, despite most FRBs being located at much further distances. This simplified treatment is motivated by the asymptotic behaviour of the lever-arm effect. Even with a conservative placement of all FRBs at 100~Mpc, the final averaged sensitivity map shows that pulse-broadening effects are subdominant to other sensitivity factors relevant to CHIME/FRB.  The mean relative sensitivity within intermediate latitudes is 99\% and effectively 100\% at high latitudes. Consequently, we should not over-interpret a lack of latitude dependence as an outright validation of the Galactic free-electron models. If we consider the other effects to be a foreground to pulse-broadening effects, we see how important modeling the beam, sky temperature, and bandwidth are. To this end, it will be worthwhile repeating the analysis for future CHIME/FRB catalogs with updated components (e.g., a sky temperature map measured directly by the CHIME cosmology experiment). Note that we have not considered any host-related scattering here, as those should not show latitude dependence and can be absorbed into the intrinsic width term.

\subsection{Completeness Cuts}\label{sec:cuts}
\begin{figure}[htb]
    \centering
    \includegraphics[width=0.95\linewidth]{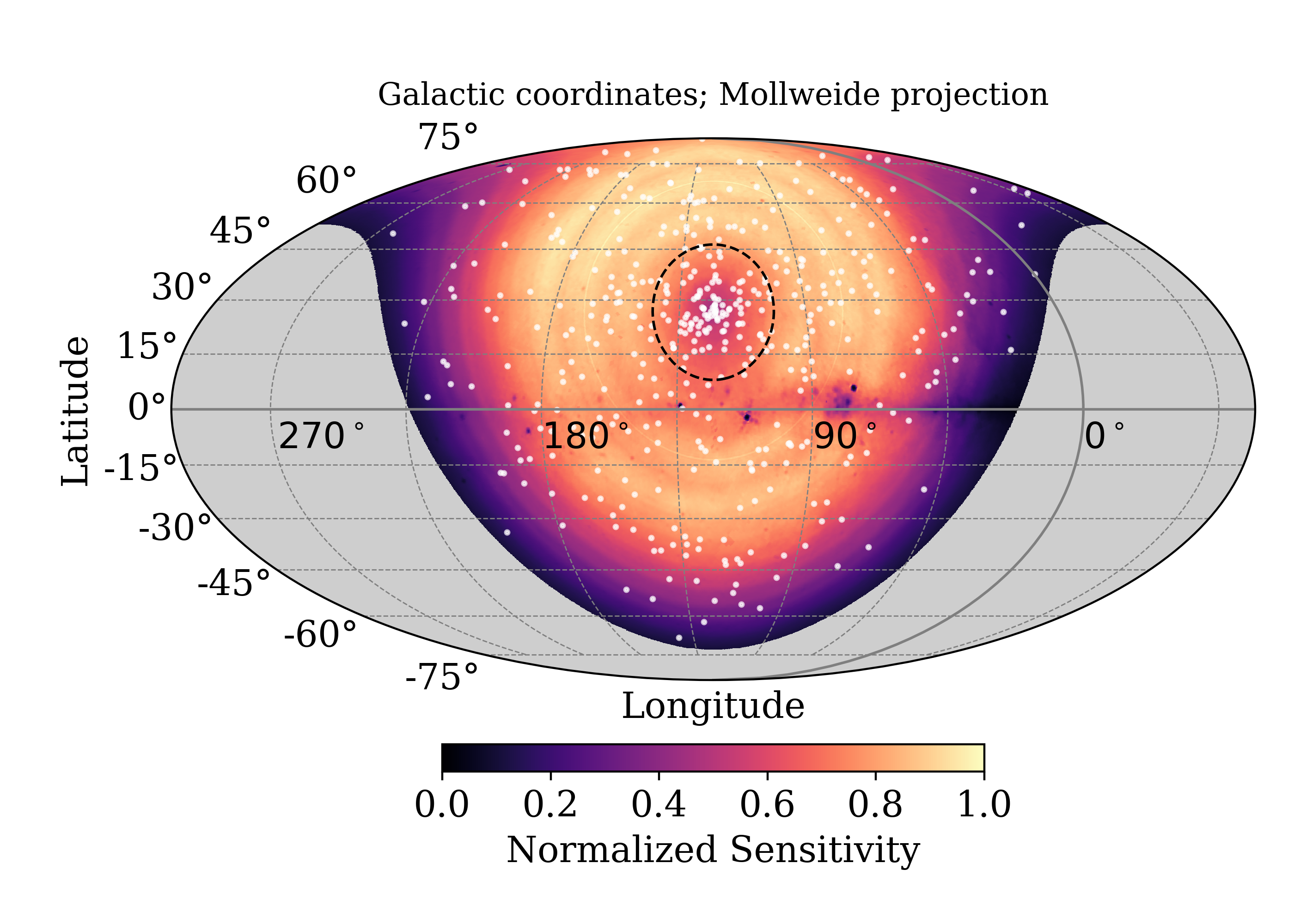}
    \caption{Normalized sensitivity for the CHIME field of view, including contributions from sky temperature, intensity beam response, masking fraction, DM intra-channel smearing, and scattering. Normalization is done with respect to the peak sensitivity. The map has been rotated 123$^\circ$ in Galactic longitude to horizontally center the North Celestial Pole. NE2001 is used here for pulse-broadening terms, with YMW16 giving a qualitatively similar map. White dots show the location of the FRBs used in this analysis. Dashed black line shows the extent of the lower transit, where exposure and sensitivity is double-valued (lower transit sensitivity not shown).  The concentric circular patterns are due to beam sensitivity (Fig.~\ref{fig:beam-profile}) and RFI masking (Fig.~\ref{fig:rfi-mask}).}
    \label{fig:sensitiviy}
\end{figure}

\begin{figure}[htb]
    \centering
    \includegraphics[width=0.95\linewidth]{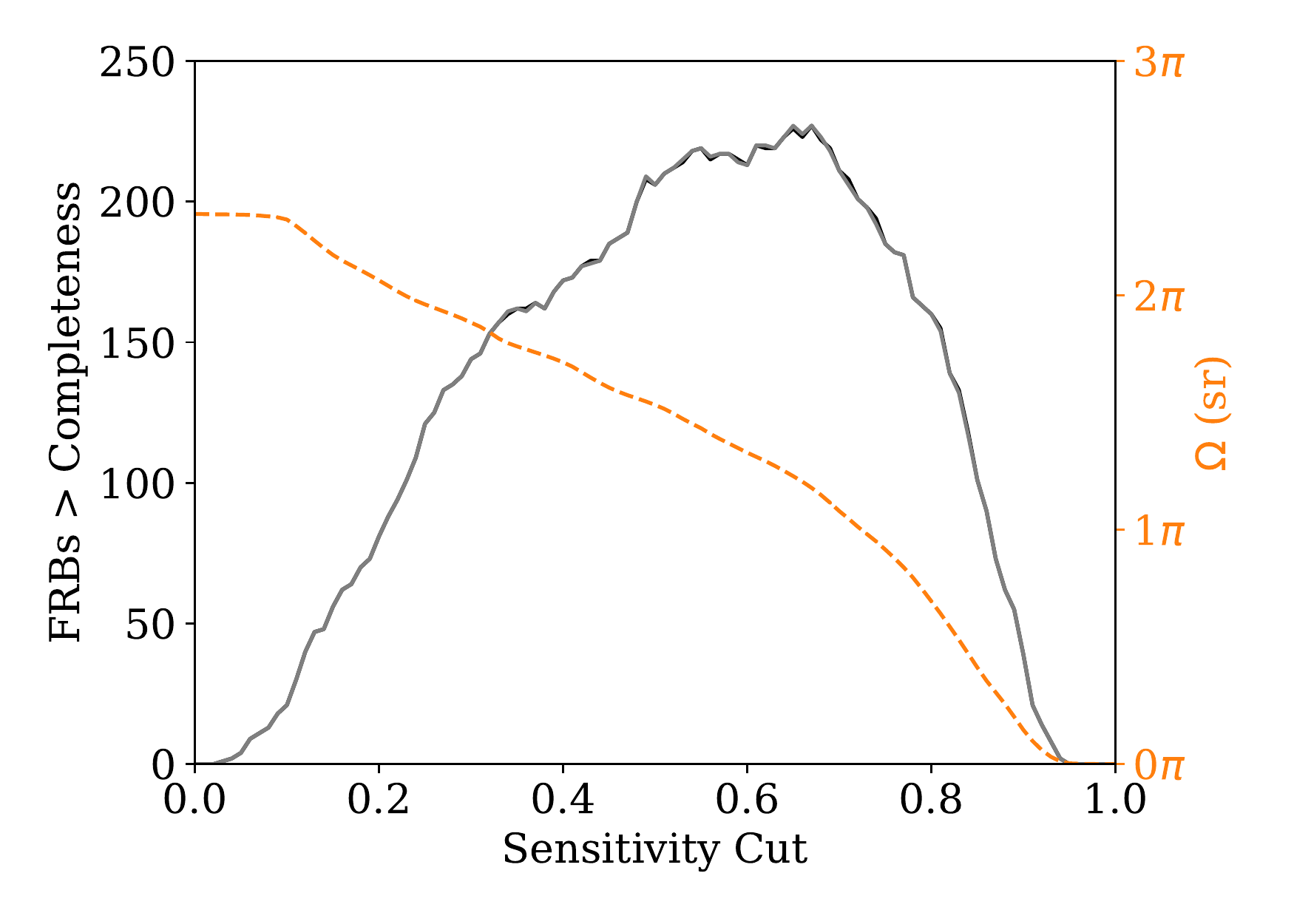}
    \caption{Number of FRBs above completeness (black solid) and sky area (orange dashed) as a function of sensitivity cutoff. As the threshold is raised, the sky area and the number of candidate FRBs for inclusion is reduced. Likewise, fewer candidates are filtered out for having insufficient S/N to survive a translation to a location with the threshold sensitivity. Whether NE2001 or YMW16 is used to calculate pulse-broadening effects, the samples sizes are nearly identical across all sensitivity cuts.}
    \label{fig:completeness}
\end{figure}

With HEALPix maps representing relative sensitivities due to differential sky temperature, effective bandwidth, forward gain, and pulse-broadening, a composite sensitivity map can be made with a simple multiplicative combination. This map is normalized to its most sensitive position (see Figure~\ref{fig:sensitiviy}).  Note that, beyond excising bursts from days with anomalous sensitivity (see \S\ref{sec:observations}), we do not incorporate any day-to-day sensitivity corrections when filtering events for completeness.   We say a burst is above completeness when we expect it to be detectable throughout the field of view, so the process of forming a completeness cut requires the sky area to be defined apriori. Rather than specifying the field of view explicitly, we make our completeness cuts by first designating a minimum sensitivity threshold (the cut). Locations in the map with a sensitivity lower than this cut are excluded from the field of view (exposure pixels are set to zero). An FRB with non-zero exposure is then considered above completeness if the following inequality is true
\begin{equation}
    \mathrm{S/N}_{\mathrm{FRB}}\times
    \frac{\mathrm{Sensitivity\ cut}}{\mathrm{Sensitivity\ at\ FRB}}\ >\ 9.
\end{equation}
A single choice for the sensitivity cut is somewhat arbitrary, although natural choices might be at half of peak sensitivity, whichever cut maximizes the sample of FRBs above completeness, or some value that balances exposure with sample size (see Figure~\ref{fig:completeness}). There are two competing effects that dictate the final sample size. First, as the sensitivity cut is increased, the field of view is decreased and the number of FRBs with non-zero exposure shrinks. Secondly, as the cut increases, the S/N scaling becomes less extreme and a larger fraction of the FRBs with non-zero exposure remain detectable. As the choice of cut is arbitrary, we repeat the statistical tests for latitude dependence for a range of sensitivity cuts.

\section{Statistical Tests}\label{sec:tests}
\subsection{Cumulative Distribution Tests}
\begin{figure*}[htb]
    \centering
    \includegraphics[width=\linewidth]{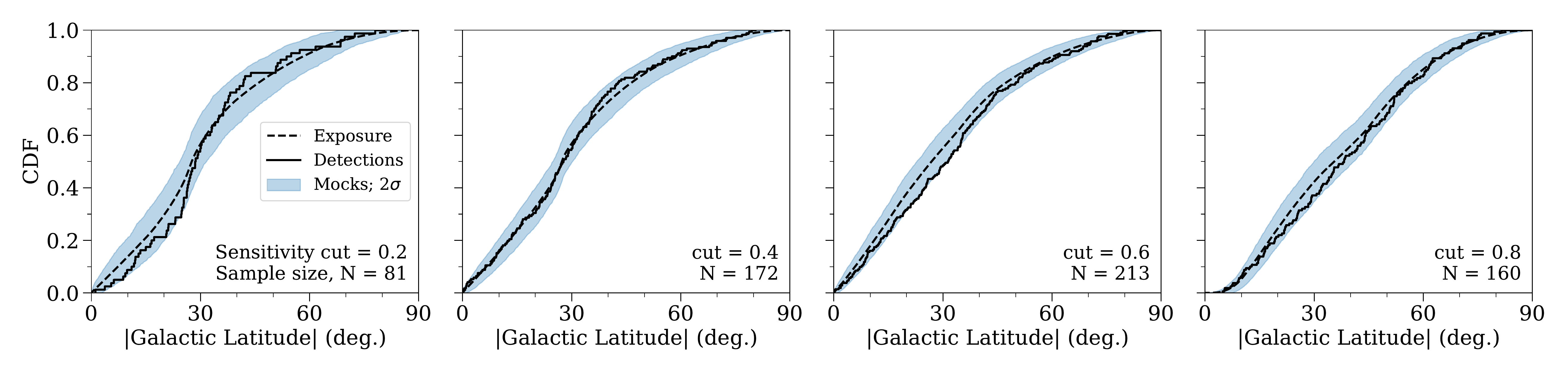}
    \includegraphics[width=\linewidth]{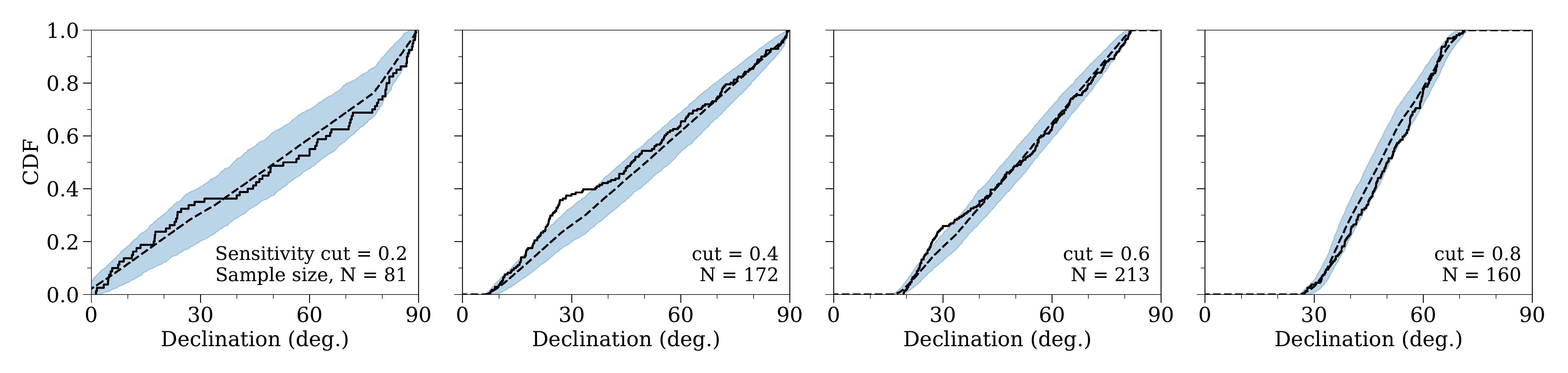}
    \caption{Distributions of detection latitudes (empirical, solid lines) and exposure (cumulative, dashed lines). Absolute Galactic latitude is shown on the top row and declination on the bottom. Each column corresponds to a different sensitivity cut and filtered sample of detections. Shaded regions are a 2$\sigma$ band derived from an ensemble of simulated detections, where each mock sample includes the same number of bursts as the observed sample and is generated from the exposure CDF.}
    \label{fig:cdf-example}
\end{figure*}

\begin{figure}[htb]
    \centering
    \includegraphics[width=0.95\linewidth]{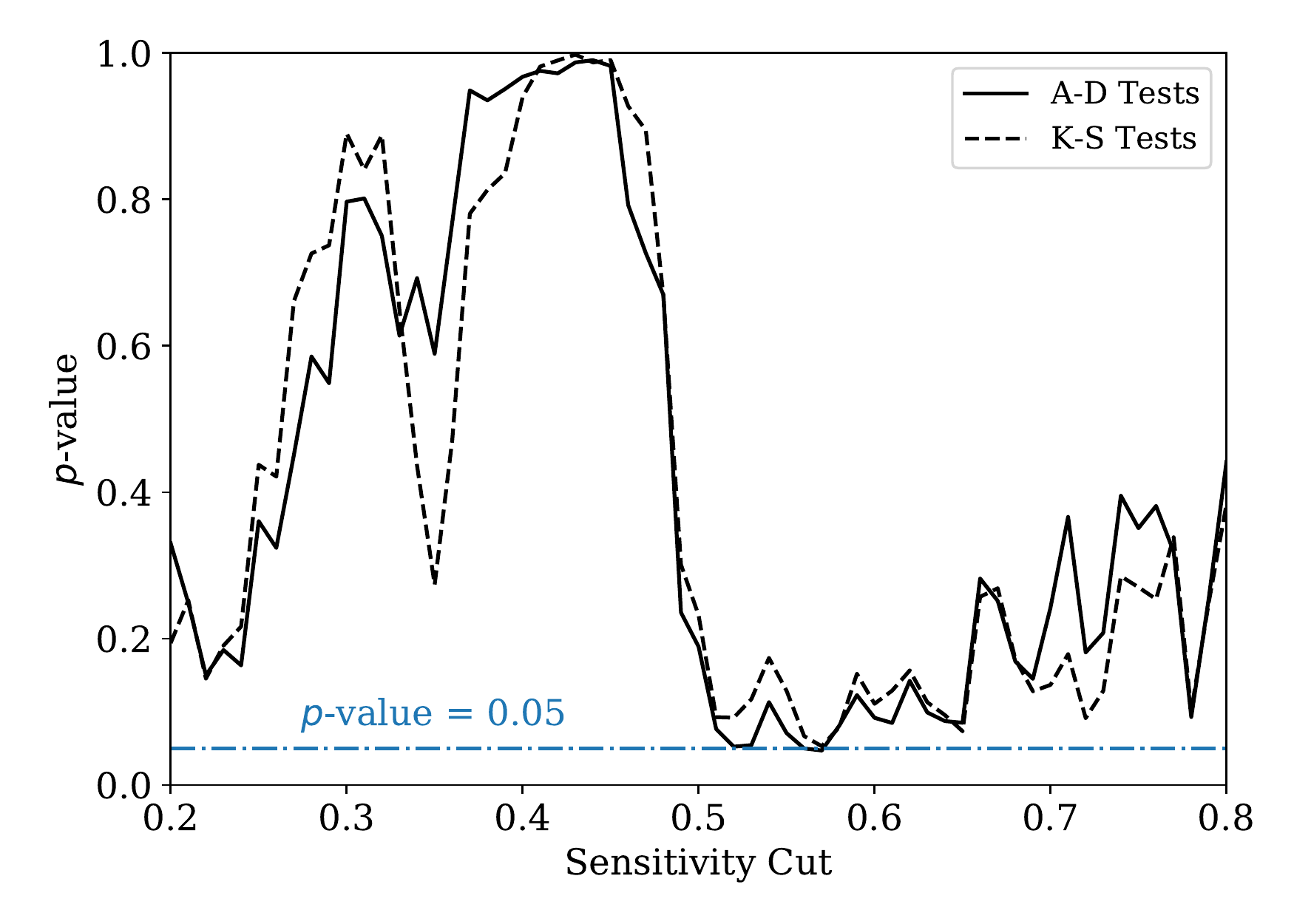}
    \caption{Anderson-Darling (solid lines) and Kolmogorov–Smirnov (dashed lines) test results for a range of sensitivity cuts. Tests compare cumulative exposure to cumulative detections (above completeness) as a function of absolute Galactic latitude.}
    \label{fig:cdf-results}
\end{figure}

Assuming sensitivity corrections have been properly handled, the detections above completeness should track exposure if the FRB rate is isotropic across the sky.  Statistical tests that ask whether an empirical distribution function (EDF) is consistent with a cumulative distribution function (CDF) are natural to apply here.  In particular, we use non-parametric one-sample Kolmogorov-Smirnov (K-S) and Anderson-Darling (A-D) tests. A-D tests are often considered preferable for certain distributions, where deviations in the tails are not realized by the difference statistic of a K-S test. Since \citet{bkb+18} concluded that a departure from isotropy was not significant via a K-S test, we include it here for comparison. The exposure CDF and detection EDF that both tests rely on are obtained as follows. After choosing a coordinate along which the comparison is to be made (e.g., declination or Galactic latitude), the reference CDF is generated from the exposure maps by weighting the coordinate value at each pixel by its exposure value. Pixels with a relative sensitivity below the chosen threshold are assigned a weight of zero. The EDF follows from the positions of the events above completeness. Examples of sensitivity corrected cumulative exposure and detection distributions are shown in Figure~\ref{fig:cdf-example} for both absolute Galactic latitude and declination, while test results for a range of sensitivity cuts are shown in Figure~\ref{fig:cdf-results}. Given the multitude of tests, their interpretation is not so straight-forward. Trial factors or look-elsewhere effects are difficult to quantify as tests performed using different sensitivity cuts are not independent (for similar thresholds, the exposure CDF and detection EDF are also  similar, hence similar $p$-values). A relevant statistic is the harmonic mean $p$-value \citep[HMP;][]{w19}, which controls for familywise error rate and, importantly, is robust to inter-dependent samples of $p$-values. For $N$ equally weighted tests, the HMP is

\begin{equation}\label{eq:hmp}
    \mathring{p} = \frac{N}{\sum_{i=1}^{N}{1/p_i}}.
\end{equation}

The HMP should only be interpreted as a false-alarm rate in the usual way for small $N$ and small $\mathring{p}$. Adjusted critical values are obtained with with the CRAN package \texttt{harmonicmeanp}\footnote{\url{https://CRAN.R-project.org/package=harmonicmeanp}}. Considering the 80 tests shown in Figure~\ref{fig:cdf-results}, the adjusted critical value for a 0.05 significance threshold is 0.037. The HMP for the A-D tests is 0.56, so we conclude the ensemble of tests does not support a Galactic latitude dependence.

\subsection{Bayesian Coin Flipping Tests}

\cite{cpo16} equate the problem of determining whether a differential rate exists to a biased coin flipping experiment and provide a Bayesian framework to test for isotropy, which we follow here. The sky is first partitioned in two, separating high-latitude and low-latitude regions about $|b| = 15^\circ$. For an isotropic sky distribution (Model 1), the probability of detecting a random burst in either region is dictated by their relative exposure. With $\alpha$ as the ratio of high-latitude exposure to low-latitude exposure, the probability of a random burst appearing in the high-latitude region is $p = \alpha/(\alpha + 1)$.  The probability of detecting $M$ out of a total $K$ events in the upper region then follows a binomial distribution

\begin{equation}
    P(M|K,p) = \binom{K}{M}p^M(1-p)^{K-M}.
\end{equation}

To conduct a Bayesian model selection, \citet{cpo16} specify an alternative possibility (Model 2), where $p$ is not defined by the relative exposure between regions, but is instead an unknown free parameter with a flat prior over $0-1$.  Assigning equal weights to each model, and marginalizing over the possible values of $p$ in the alternative model, the ratio of posterior probabilities (Bayes factor) for the two models is 

\begin{equation}
    \frac{P(M|\mathrm{Model\ 1}, K)}{P(M|\mathrm{Model\ 2}, K)} = (K+1)\binom{K}{M}p_1^M(1-p_1)^{K-M}.
\end{equation}

\begin{figure}[htb]
    \centering
    \includegraphics[width=0.95\linewidth]{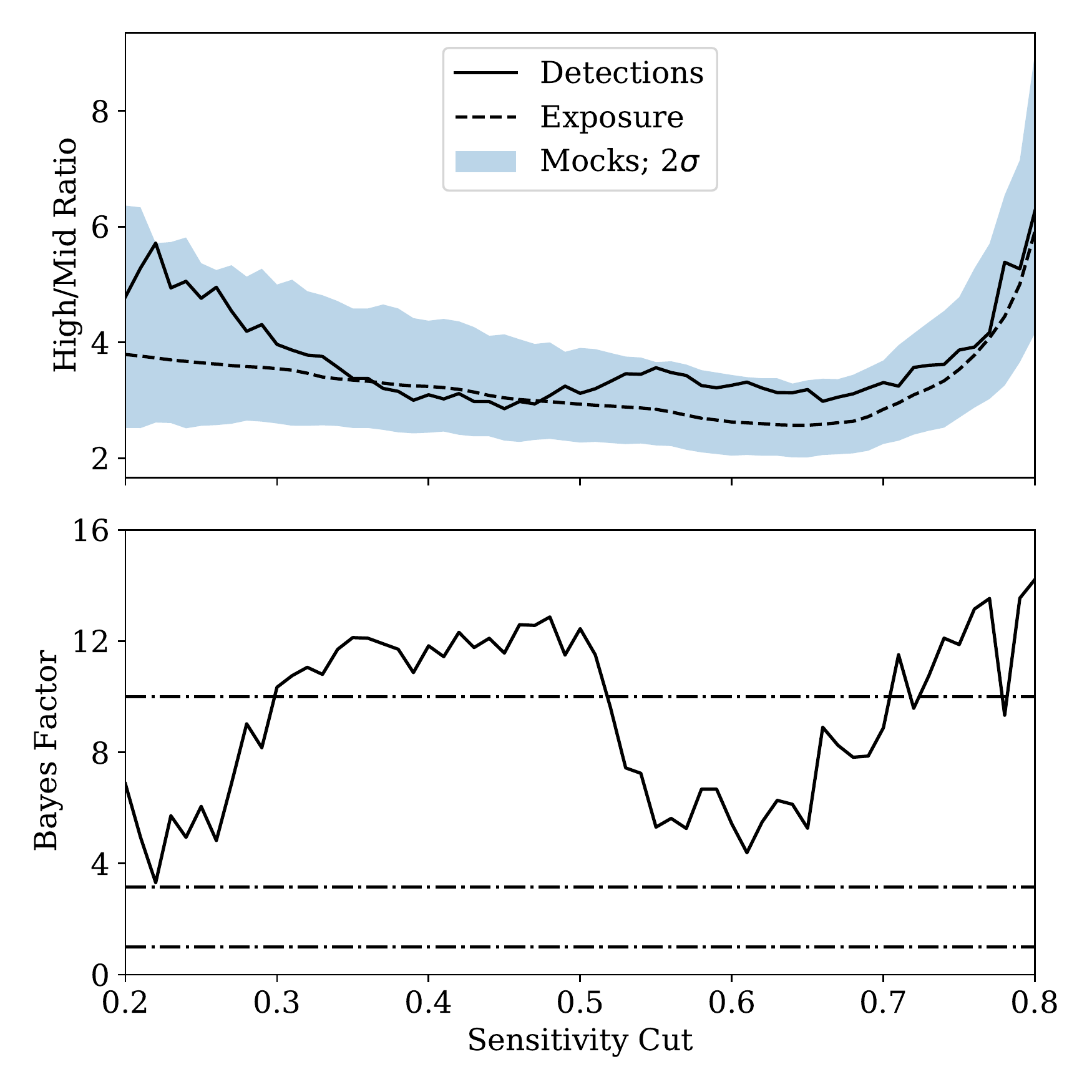}
    \caption{Diagnostic plots for Bayesian biased coin flip tests performed for a variety of sensitivity cuts. Upper panel: the sky is partitioned into high and intermediate latitudes at $|$b$|$ = 15$^\circ$ -- ratios between these regions are shown for both the detections (solid line) and the exposure (dashed line). The shaded region is a 2$\sigma$ band derived from an ensemble of simulated ``coin flips", where each mock includes the same number of events as the observed sample and is generated from a binomial distribution with bias specified by the exposure ratio. Lower panel: Associated Bayes factor for each sensitivity cut (solid line), along with common significance thresholds (dashed lines), with ``strong" evidence for isotropy at a Bayes factor of 10, ``significant" at 10$^{1/2}$, and ``barely worth mentioning" at 1.}
    \label{fig:coin-results}
\end{figure}

Once again we perform this model selection test for a range of sensitivity cuts, where each cut defines the values of $K$, $M$, and $\alpha$ (which gives $p_1$). Resulting Bayes factors along with the exposure ratio $\alpha$ and detection ratio $K/(M-K)$ are shown in Figure~\ref{fig:coin-results}. To interpret the significance of the resulting Bayes factors, we use the scale suggested by \citet{kr95}, where the evidence for Model 1 over Model 2 is ``strong" when the Bayes factor is greater than 10, ``substantial" when between 10$^{1/2}$ and 10, and ``barely worth mentioning" between 1 and 10$^{1/2}$. We find the evidence for isotropy is strong for roughly half of the sensitivity cuts, and substantial for the remaining half.

\subsection{Cross-checks}
Given the large size of the initial sample of CHIME/FRB events, we are afforded the luxury of performing a number of additional cross-checks that would be difficult with previous FRB samples. We can check the robustness or stability of results by jackknife resampling, and we can partition the sample with respect to a variety of parameters to check for possible latitude deficits within subsections of the parameter-space. For example, intrinsically narrow bursts may suffer from Galactic scatter-broadening or higher frequency bursts may be suppressed through scintillation effects. Test results for sample splits are summarized in Table~\ref{tab:splits}. Of note here is the upper excess DM split (DM$_X >$ median DM$_X$), where the HMP is below the adjusted critical value for 0.05 significance. When compared to initial claims, the significance is somewhat marginal (false-alarm rate of ${\sim}$2\%). It is not clear why events with higher excess DM near the plane are under-represented in our sample and it will be interesting to revisit this analysis with future CHIME/FRB catalogs to see if the effect persists. 

We have also investigated splitting the sample into one-offs and repeaters. The results are effectively the same with respect to the A-D tests, while for the Bayes tests, the previously strong evidence for isotropy is diminished to being ``barely worth mentioning".  However, we caution that the sample size of repeaters used in this analysis is quite small in comparison to the one-offs. With only order 10 sources surviving the sensitivity cuts, the current statistical power is quite borderline. CHIME/FRB will continue to expand the pool of repeaters, and this analysis should be revisited as the population grows.

A-D test results along Galactic and celestial coordinates are summarized in Table~\ref{tab:coordinates}, including tests against both absolute and signed Galactic latitude. Elsewhere, we have chosen to focus on absolute latitude by default, under the assumption that there is no sign dependence, and that a more dense sampling of the parameter space is statistically preferable. In both cases the HMP is unremarkable, suggesting our symmetry assumption is valid. Deviations from isotropy with respect to declination are indicative of unmodeled instrumental effects, so the excess of events just below declination $\delta\sim30^\circ$ is cause for concern (see Figure~\ref{fig:cdf-example}, bottom row). For declination, the A-D $p$-value is minimal (0.01) at a sensitivity cut of 0.32. The HMP is above the adjusted critical value for a 0.05 significance, so we say the ensemble of declination tests does not reject isotropy. Nevertheless, we investigate the worst-case sensitivity cut by taking half of the post-cut events randomly (without replacement) and repeating the A-D test 1000 times. The mean $p$-value of these jackknife resampling tests is 0.12, suggesting the deviation from isotropy is marginal. Furthermore, splitting the sample on arrival time, the latter half does not show a statistically significant excess. It is not clear what would explain the excess-- events in that declination range are not clustered in right ascension or DM, so a missed repeater does not work as an explanation for the excess.  An increased sensitivity may be expected for declinations close to the sources responsible for the daily system-wide gain calibration. These sources: Cygnus A ($\delta \sim +41^\circ$), Cassiopeia A ($\delta \sim +59^\circ$), Taurus A ($\delta \sim +22^\circ$), and Virgo A ($\delta \sim +12^\circ$), are cycled depending on when their transits occur.  During the first half of Catalog 1 Taurus A was not used, so an increased sensitivity in that declination regime is not expected.  Repeating the A-D tests for Galactic latitude using only the latter half, the $p$-values remain insignificant.
As these methods are sensitive to both scarcity and excess, it is reasonable to investigate questions of possible Galactic contamination. A concrete limit would be relevant to evaluations of NE2001 and YMW16, and would be a useful figure of merit for any low DM population inferences of Catalog 1. Fully quantifying what fraction of the sample could be of Galactic origin is beyond the scope of this paper, as to do so accurately would require integrating the CHIME/FRB selection function with population synthesis models.  However, working towards a rough approximation, we have considered a simple limiting case where selection effects are ignored, and the contaminating population is composed entirely of young isolated neutron stars.  This scenario concentrates potential sources to low Galactic latitude, which should have the strongest statistical impact. We first simulate a large population of fake sources following the prescription from \citet{fk06}.  After projecting onto the sky, this population can be randomly sampled using the exposure maps as weights.  We then repeated the A-D tests shown in Figure 7, progressively increasing the number of fake sources we add to the real sample.  We find the HMP passes the critical points equivalent to 5\%, 1\%, and 0.1\% false-alarm rates, when the contaminated fraction of the final sample reaches roughly 7\%, 8\%, and 10\% respectively. Again, this should only be taken as a crude estimate of how sensitive these methods are to varying degrees of Galactic contamination. A full analysis should incorporate priors on allowable DM excess and impose selection effects on the synthesized population, which should include evolved pulsars and RRATs.

\begin{table}[]
    \centering
    \begin{tabular*}{\columnwidth}{@{\hspace{3em}}r@{\hspace{0.5em}}l@{\hspace{3em}}ccc}
        \addlinespace[1em]
        \toprule
        & & \multicolumn{3}{c}{\makebox[0pt]{A-D $p$-value}} \\
        \cmidrule(r){3-5}
        & & min & HMP & max  \\
        \midrule
        Galactic & Longitude & 0.40 & 0.64 & 0.97 \\
         & Latitude & 0.02 & 0.23 & 0.77 \\
         & $|$Latitude$|$ & 0.05 & 0.18 & 0.99 \\ \addlinespace[1em]
         \multicolumn{2}{c}{Right Ascension} & 0.11 & 0.48 & 0.98 \\
         \multicolumn{2}{c}{Declination} & 0.01 & 0.07 & 0.48 \\ 
        \bottomrule
    \end{tabular*}
    \caption{Min, harmonic mean (Eqn.~\ref{eq:hmp}), and max $p$-values from Anderson-Darling tests performed over a range of sensitivity cuts from 0.2 to 0.8. Each batch of tests compares cumulative detections with the cumulative exposure as a function of different coordinates. In none of the coordinate frames is the null hypothesis of isotropy consistently or strongly rejected. Note that the critical $p$-value of 0.05 for significance is adjusted to 0.037 for the HMP test. K\nobreakdash-S test results are similar in all cases and are therefore not shown (see Figure~\ref{fig:cdf-results}).}
    \label{tab:coordinates}
\end{table}

\begin{table}[]
    \centering
    \begin{tabular*}{0.95\columnwidth}{rl@{\hspace{1.5em}}ccc@{\hspace{1.5em}}ccc}
        \addlinespace[1em]
        \toprule
        & & \multicolumn{3}{c}{A-D $p$-value\hphantom{1em}} & \multicolumn{3}{c}{Bayes factor} \\
        \cmidrule(r{1.5em}){3-5} \cmidrule(r{0.5em}){6-8}
        & & min & HMP & max & min & mean & max\\
        \midrule
        \multicolumn{2}{c}{No split} & 0.05 & 0.18 & 0.99 & 3.3 & 9.3 & 14.2 \\ \addlinespace[1em]
        \multicolumn{2}{c}{One-offs} & 0.05 & 0.17 & 0.98 & 2.5 & 8.9 & 13.9 \\
        \multicolumn{2}{c}{Repeaters} & 0.09 & 0.25 & 0.89 & 0.4 & 2.3 & 3.5 \\
        \addlinespace[1em]
        \multirow{2}{*}{DM$_{X}$} & $\uparrow$ & 0.002 & 0.02 & 0.33 & 0.5 & 2.2 & 7.2 \\
        & $\downarrow$ & 0.10 & 0.33 & 0.87 & 1.5 & 6.4 & 9.5 \\ \addlinespace[1em]
        \multirow{2}{*}{$\tau_{sc}$} & $\uparrow$& 0.17 & 0.48 & 0.96 & 3.5 & 6.8 & 9.2 \\
        & $\downarrow$ & 0.01 & 0.08 & 0.90 & 2.0 & 7.1 & 9.8 \\ \addlinespace[1em]
        \multirow{2}{*}{$w_i$} & $\uparrow$& 0.01 & 0.07 & 0.73 & 1.4 & 5.3 & 9.0 \\
        & $\downarrow$ & 0.04 & 0.20 & 0.99 & 1.4 & 6.5 & 9.6 \\ \addlinespace[1em]
        \multirow{2}{*}{$\nu_c$} & $\uparrow$& 0.34 & 0.63 & 0.99 & 3.4 & 7.3 & 9.4 \\
        & $\downarrow$ & 0.01 & 0.07 & 0.98 & 2.3 & 5.6 & 10.7 \\
        \bottomrule
    \end{tabular*}
    \caption{Test results for (absolute) Galactic latitude dependence for different splits of the initial sample, where up/down arrows denote upper/lower splits, partitioned at the median value. The min/(harmonic)mean/max values summarize a range of tests with varying sensitivity threshold (from 0.2 to 0.8 of peak sensitivity).  DM, $\tau_{sc}$, and $w_i$ are relevant for pulse-broadening sensitivity corrections, while $\nu_c$ (peak frequency of spectral fit) has relevance for frequency dependent effects such as scintillation. Note that the critical $p$-value of 0.05 for significance is adjusted to 0.037 for the HMP test.}
    \label{tab:splits}
\end{table}

\section{Discussion}
\label{sec:discussion}

Using 453 events from Catalog 1, we have searched for evidence of dependence of the FRB sky distribution on Galactic latitude, but have found none, as expected if FRBs are a cosmological population and selection effects have been sufficiently accounted for.
This is in contrast to past reports
\citep{psj+14,bb14, cpk+16} which found avoidance of the Galactic Plane in surveys conducted at 1.4~GHz, but which were based on far fewer events than considered here.

\citet{mj15} considered the previously reported Plane avoidance and suggested that the effect was due to diffractive scintillation within the Milky Way, with the underlying FRB sky distribution being isotropic.
The idea was that an effect associated with Eddington bias artificially enhanced the high Galactic latitude rate, with the enhancement expected to be greater for a steeper FRB flux distribution.

The FRB sky isotropy we report for Catalog 1 events may not, however, shed light on the effect considered by \citet{mj15},
who argued an enhancement at high latitudes was plausible if the scattering imposed by the Galaxy was substantially overestimated by electron density models, and extragalactic scattering was not so extreme (e.g., $\tau_{sc} < 1$ ms at 1.4~GHz) as to alter the coherence properties and quench interstellar scintillation. With the basic requirement of having a decorrelation bandwidth $\Delta\nu_d$ comparable or larger than roughly half the observing bandwidth, their argument was supported by the existence of several high-latitude pulsars with $\Delta\nu_d \sim 300$~MHz at 1.4~GHz in conjunction with the ${\sim}$300~MHz bandwidth used in early FRB detections {with the Murriyang telescope}. While these observed decorrelation bandwidths greatly exceed modeled expectations, they are strongly diminished when scaled to CHIME frequencies (400--800 MHz), suggesting the effect would be unimportant in the CHIME band.

A Galactic latitude dependence may also be expected if there are systematic errors in Milky Way electron density distribution models \citep{ne2001,ymw17}.
These could lead to Galactic RRATs and radio pulsars, predominantly a Plane population, misidentified as extragalactic FRBs.  Specifically, underestimates of the Galactic DM contribution in the Plane would preferentially result in higher Plane FRB rates, and the converse. That we find no latitude dependence implies any inaccuracies in the DM models are not strong enough to meaningfully impact the CHIME/FRB detection rate as a function of latitude. Given the subdominance of pulse-broadening effects (see \S\ref{sec:pulse-broadening}), we caution that these limits are not as stringent as one might hope. The lack of latitude dependence is nevertheless interesting given that these models are calibrated very differently at high Galactic latitude versus low. For example, of the 189 independent distance estimates used by \citet{ymw17} to calibrate their Galactic DM model, for $|b|<5^{\circ}$, 77\% are from either \ion{H}{1} kinematics measurements or associated nebulae, while these two types are used for only 4\% of calibrators at $|b|>5^{\circ}$.  Our results therefore suggest that either their calibration methods are as reliable as they believe, that any problems with methods fortuitously cancel, or that the current sample size and foreground modeling for CHIME/FRB is insufficient (given the relative insensitivity to pulse-broadening effects).  A lack of evidence for significant systematics in these models suggests that efforts to use FRBs to constrain the DM content of the Galactic halo are worthwhile.

\section{Conclusions}\label{sec:conclusions}

In this work we have taken a large subset of events from the first CHIME/FRB catalog, occurring during relatively uniform observing conditions, with the goal of determining whether a Galactic latitude dependence exists in the FRB sky distribution. We constructed a radiometer-based relative sensitivity map over the CHIME/FRB field of view which accounts for sky temperature, bandwidth, intensity beam response, and pulse-broadening effects. With this map, we choose a threshold, which sets the field of view and allows us to filter out events considered below completeness.
We performed one-sample Kolmogorov-Smirnov and Anderson-Darling tests over a range of sensitivity cuts, but find no evidence to reject the null hypothesis of isotropy with respect to Galactic latitude.  We then partition the sky into high and intermediate latitude regions and perform a Bayesian model selection based on a biased coin flipping treatment, concluding the weight of evidence for isotropy ranges from substantial to strong (depending on the sensitivity threshold).  Finally, we explored a variety of cross-checks but find no strong evidence for any deviation from isotropy. Our results are thus consistent with FRBs being a cosmological population, and we find no evidence for differential systematic errors in Galactic DM models, in spite of them having calibration strategies that depend strongly on latitude.

\begin{acknowledgements}
We acknowledge that CHIME is located on the traditional, ancestral, and unceded territory of the Syilx/Okanagan people.

We thank Emily Petroff for discussions that have improved the quality of this manuscript.

We thank the Dominion Radio Astrophysical Observatory, operated by the National Research Council Canada, for gracious hospitality and expertise. CHIME is funded by a grant from the Canada Foundation for Innovation (CFI) 2012 Leading Edge Fund (Project 31170) and by contributions from the provinces of British Columbia, Qu\'ebec and Ontario. The CHIME/FRB Project is funded by a grant from the CFI 2015 Innovation Fund (Project 33213) and by contributions from the provinces of British Columbia and Qu\'ebec, and by the Dunlap Institute for Astronomy and Astrophysics at the University of Toronto. The Dunlap Institute is funded through an endowment established by the David Dunlap family and the University of Toronto. Additional support was provided by the Canadian Institute for Advanced Research (CIFAR), McGill University and the McGill Space Institute via the Trottier Family Foundation, and the University of British Columbia. Research at Perimeter Institute is supported by the Government of Canada through Industry Canada and by the Province of Ontario through the Ministry of Research \& Innovation. The National Radio Astronomy Observatory is a facility of the National Science Foundation (NSF) operated under cooperative agreement by Associated Universities, Inc. FRB research at UBC is supported by an NSERC Discovery Grant and by the Canadian Institute for Advanced Research. The CHIME/FRB baseband system is funded in part by a CFI John R. Evans Leaders Fund award to I.H.S..
FRB research at UBC is supported by an NSERC Discovery Grant and by the Canadian Institute for Advanced Research.
D.M. is a Banting Fellow.
J.M.P. is a Kavli Fellow.
K.S. is supported by the NSF Graduate Research Fellowship Program.
K.W.M. is supported by an NSF Grant (2008031).
M.B. is supported by an FRQNT Doctoral Research Award.
P.C. is supported by an FRQNT Doctoral Research Award.
P.S. is a Dunlap Fellow and an NSERC Postdoctoral Fellow. 
B.M.G. acknowledges the support of the Natural Sciences and Engineering Research Council of Canada (NSERC) through grant RGPIN-2015-05948, and of the Canada Research Chairs program.
V.M.K. holds the Lorne Trottier Chair in Astrophysics \& Cosmology, a Distinguished James McGill Professorship and receives support from an NSERC Discovery Grant (RGPIN 228738-13) and Gerhard Herzberg Award, from an R.~Howard Webster Foundation Fellowship from CIFAR, and from the FRQNT CRAQ.

\end{acknowledgements}

\bibliography{frbrefs}{}
\bibliographystyle{aasjournal}

\end{document}